\documentclass[aps,prl,twocolumn,nofootinbib,superscriptaddress]{revtex4}
\usepackage{graphicx}
\usepackage{times}
\usepackage{epsfig}
\usepackage{amsmath}
\usepackage{amsfonts}
\usepackage{amssymb}
\usepackage{url}
\usepackage{hyperref}
\usepackage{subfigure}
\newcommand{\be}{\begin{equation}}
\newcommand{\ee}{\end{equation}}
\newcommand{\bea}{\begin{eqnarray}}
\newcommand{\eea}{\end{eqnarray}}

\begin{document}

\pagestyle{plain}

\title{
 Diquark Higgs at LHC
}

\author{R. N. Mohapatra\footnote{
  e-mail: rmohapat@physics.umd.edu},
        Nobuchika Okada\footnote{
  e-mail: okadan@post.kek.jp, on leave of absence from
          Theory Division, KEK, Tsukuba 305-0801, Japan},
    and Hai-Bo Yu\footnote{
  e-mail: hbyu@physics.umd.edu}}

\affiliation{Department of Physics,
 University of Maryland, College Park, MD 20742, USA}

\date{September, 2007}

\preprint{\vbox{\hbox{UMD-PP-07-010}}}

\bigskip
\bigskip

\begin{abstract}
Existence of color sextet diquark Higgs fields with TeV masses
will indicate a fundamentally different direction for unification
than conventional grand unified theories. There is a class of
partial unification models based on the gauge group $SU(2)_L\times
SU(2)_R\times SU(4)_c$ that implement the seesaw mechanism for
neutrino mass with seesaw scale around $10^{11}$ GeV, where indeed
such light fields appear naturally despite the high gauge symmetry
breaking scale. They couple only to up-type quarks in this model.
 We discuss phenomenological constraints on these fields and show that
 they could be detected at LHC via their decay to either $tt$ or
single top + jet. We also find that existing Tevatron data gives a
lower bound on its mass somewhere in the 400-500 GeV, for
reasonable values of its coupling.
\end{abstract}
\maketitle

{\bf Introduction:}
The Large Hadron Collider (LHC), starting its operation within a
year is expected to probe a new hitherto unexplored domain of
particles and forces beyond the standard model. It can not only
clarify some of the many mysteries of the standard model but also
perhaps provide a glimpse of other new physics in the TeV energy
range. The sense of expectation generated by this in the particle
physics community has led to a burst of theoretical activity
designed to explore great many theoretical concepts that perhaps the
LHC can throw light on. They include ideas such as extra dimensions,
supersymmetry, new strong forces, new Higgs bosons, new quarks and
leptons etc. In this paper, we explore the possibility that LHC can
throw light on a new kind of color sextet Higgs fields (denoted by
$\Delta_{u^cu^c}$). Existence of such fields will indicate a
fundamentally new direction for unification than the conventional
grand unified theories. Indeed, there is a class of partial
unification theories based on supersymmetric $SU(2)_L\times
SU(2)_R\times SU(4)_c$ \cite{ps} model where $\Delta_{u^cu^c}$
fields appear with mass in the TeV range even though the gauge
symmetry breaking scale is in the range of $10^{11}$ GeV due to the
existence of accidental symmetries \cite{chacko}. These models are
interesting since they not only unify quarks and leptons but also
implement the seesaw mechanism for small neutrino masses
\cite{seesaw} and therefore provide a theoretical basis to
contemplate the existence of TeV scale diquark Higgs fields. They
make the unique prediction that $\Delta_{u^cu^c}$ fields couple only
to the right-handed up-type quarks of all generations. These
particles are also connected to baryon number violating processes
such as neutron-anti-neutron oscillation \cite{chacko}.
Discovery of these  particles would point towards quark-lepton
unification at an intermediate scale rather than at the commonly
assumed grand unification scale of $10^{16}$ GeV.

An interesting point about these particles is that they can be
produced and detected at the LHC. Their couplings are however
constrained by low energy observations. In this letter we explore
this topic and the main results of our investigation are:
\begin{itemize}
\item the present  experimental  information on
$D^0-\overline{D^0}$ mixing can be satisfied by setting to zero
only the diagonal coupling of the $\Delta_{u^cu^c}$ to the charm
quarks; \item the remaining coupling can be large enough so that
the production rate in $pp$ collision is significant and there are
observable signal for the diquark Higgs field via its double top
and single top plus jet production. Note also that a $pp$
colliding machine such as LHC is more favorable for the production
of these kind of fields compared to a $p\overline{p}$ machine e.g.
Tevatron; \item the $\Delta_{u^cu^c}$ coupling matrix to quarks
can be a direct measure of the neutrino mass matrix if the
neutrino masses have a inverted hierarchy within our scheme,
providing a unique way to probe the lepton flavor structure using
the LHC.
\end{itemize}

%
%

{\bf Brief overview of model with naturally light
$\Delta_{u^cu^c}$: }
In order to theoretically motivate our study of color sextet Higgs
fields, we discuss how these ``light mass'' particles can naturally
arise in a class of supersymmetric seesaw models for neutrino masses
\cite{seesaw}. The seesaw mechanism extends the standard model with
three right handed neutrinos and add large Majorana masses for them.
 The fact that the seesaw scale is much
lower than the Planck scale suggests that there may be a symmetry
protecting this scale. A natural symmetry is local B-L symmetry
whose breaking leads to the right-handed Majorana neutrino masses.
A gauge theory that accommodates this scenario is the left-right
symmetric model based on the gauge group $SU(2)_L\times
SU(2)_R\times U(1)_{B-L}\times SU(3)_c$. This model being quark
lepton symmetric easily lends itself to quark-lepton unification a
la Pati-Salam into the gauge group
 $SU(2)_L\times SU(2)_R\times SU(4)_c$ \cite{ps}. It has already
 been shown \cite{chacko} that
within a supersymmetric Pati-Salam scheme, if $SU(4)_c$ color is
broken not by $SU(2)_{L,R}$ doublet fields as was suggested in
 \cite{ps} but rather by triplets as proposed in \cite{goran}, then
despite the high seesaw scale of around $10^{11}$ GeV or so, there
are light (TeV mass) sextet diquark of the type $\Delta_{u^cu^c}$.

To show this more explicitly, recall that the quarks and leptons in
this model are unified and transform as $\psi:({\bf 2,1,4})\oplus
\psi^c:({\bf 1,2},\overline{\bf 4})$ representations of
$SU(2)_L\times SU(2)_R\times SU(4)_c$. For the Higgs sector, we
choose, $\phi_1:(\bf{2,2,1})$ and $\phi_{15}:(\bf{2,2,15})$ to give
mass to the fermions and the $\Delta^c:({\bf 1,3,10})\oplus
\overline{\Delta}^c:({\bf 1,3},\overline{\bf 10})$ to break the
$B-L$ symmetry. The diquarks mentioned above are contained in the
$\Delta^c:(\bf{1,3,10})$ multiplet.

The renormalizable superpotential for this model has a large
global symmetry of $U(30,c)$ and on gauge symmetry breaking, leads
to all diquark Higgs fields and a pair of doubly charged Higgs
bosons remaining light. In this theory, the gauge couplings become
non-perturbative in the $10-100$ TeV range and do not yield a high
seesaw scale, as may be desirable. On the other hand, if we add an
extra $B-L$ neutral triplet Higgs field $\Omega:(\bf{1,3,1})$ to
this theory, the symmetry of the theory gets lowered and this
helps to greatly reduce the number of light diquark states. The
reduction of the global symmetry can be seen from the
superpotential of this model $ W~=~W_Y~+~W_H$
where
\begin{eqnarray}
W_H&=&\lambda_1 S( \Delta^c\overline{\Delta}^c-M_\Delta^2)
 +\mu_{i}{\rm Tr}\,(\phi_i\phi_i) \,,\\
W_Y&=&h_1\psi\phi_1 \psi^c + h_{15} \psi\phi_{15} \psi^c + f
\psi^c\Delta^c \psi^c.
 \label{Yukawa}
\end{eqnarray}
Note that since we do not have parity symmetry in the model, the
Yukawa couplings $h_1$ and $h_{15}$ need not symmetric matrices.
This superpotential has $U(10,c)\times SU(2)$ global symmetry.
When the neutral component of $(\bf{1,3,10+\overline{10}})$ picks
up VEV, this symmetry breaks down to $U(9,c)\times U(1)$, leaving
21 complex massless scalar fields. Since the gauge symmetry also
breaks down from $SU(2)_R\times SU(4)_c$ to $SU(3)_c\times
U(1)_Y$, nine of these are absorbed leaving 12 complex massless
states, which are the sextet $\Delta_{u^cu^c}$ (the submutiplet of
the $\Delta^c$ in Eq. (\ref{Yukawa})) plus its complex conjugate
states from the ${\bf \overline{10}}$ representation above. Once
supersymmetry breaking effects are included and higher dimensional
terms
\begin{eqnarray}
&& \lambda_A \frac{(\Delta^c\overline{\Delta}^c)^2}{M_{P\!\ell}}
 +\lambda_B\frac{(\Delta^c{\Delta^c})(\overline{\Delta}^c\overline{\Delta}^c)}{
M_{P\!\ell}}  \nonumber \\
&+& \lambda_C \Delta^c\overline{\Delta}^c\Omega
 + \lambda_D \frac{{\rm Tr}\,(\phi_1\Delta^c
\overline{\Delta}^c\phi_{15})}{M_{P\!\ell}} \, ,
\end{eqnarray}
are included, these  $\Delta_{u^cu^c}$ fields pick up mass of order
$\lambda_B\frac{v^2_{BL}}{M_{Pl}}$ which for $v_{BL}\sim 10^{11}$ GeV is
in
the 100 GeV to TeV range naturally.
We denote the mass of $\Delta_{u^cu^c}$ by $m_\Delta$.

{\bf Phenomenological constraints on $\Delta_{u^cu^c}$ couplings to
quarks:}
The magnitudes of the couplings of diquark Higgs to up-type quarks
are important for its LHC signal as well as other manifestations
in the domain of rare processes. As is clear from Eq.~(\ref{Yukawa}),
 the sextet $\Delta_{u^cu^c}$ couplings to quarks, $f_{ij}$ are also
directly related to the neutrino masses, which provides a way to
probe neutrino masses from LHC observations. Due to the existence
of other parameters, current neutrino observations do
not precisely pin down the $f_{ij}$. There are however other
constraints on them.

To study these constraints, we define the $\Delta_{u^cu^c}$
couplings ($f_{ij}$) in a basis where the up-type quarks are mass
eigenstates. A major constraint on them comes from the
$D^0-\overline{D^0}$ mixing which is caused by the exchange of
$\Delta_{u^cu^c}$ field:
\begin{eqnarray}
M_{D^0-\overline{D^0}}~=~\frac{f_{11}f_{22}}{4m^2_{\Delta}}\overline{c}\gamma_\mu
(1-\gamma_5)u \overline{c}\gamma^\mu(1-\gamma_5) u ;
\end{eqnarray}
The present observations \cite{ddbar} imply that the transition
mass $\Delta M_D$ for $D^0-\overline{D^0}$ to be $8.5\times
10^{-15} \leq \Delta M_D\leq 1.9\times 10^{-14}$ in GeV units. In
our model, we can estimate this to be
\begin{eqnarray}
\Delta M_D\simeq \frac{f_{11}f_{22}}{4m^2_{\Delta}}f^2_DM_D
\end{eqnarray}
which implies that $\frac{f_{11}f_{22}}{4m^2_{\Delta}}\leq 10^{-12}$
GeV$^{-2}$; for a TeV delta mass, which is in the range of our
interest, this implies $f_{11}f_{22}\leq 4\times 10^{- 6}$. If we assume
that $f_{11}\gg f_{22}$, then for $f_{11}\sim 0.1$ or so, $f_{22}$
is close to zero, which assume to be the case in our
phenomenological analysis \cite{pakvasa}.

Next constraint comes from non-strange pion decays e.g.
$D\rightarrow \pi\pi$ which are suppressed compared to the decays
with strange final states. This bound however is weak. The present
limits on such non-strange final states are at the level of $B\leq
10^{-4}$ \cite{PDG},  which implies $f_{11}f_{12}\leq 4\times
10^{-2}$ for $m_\Delta \sim$ few hundred GeV to TeV range. This
will be easily satisfied if $f_{11}\sim f_{12}\sim 0.2$.

{\bf Collider phenomenology:}
Due to the diquark Higgs coupling to a pair of up-type quarks, it
can be produced at high energy hadron colliders
 such as Tevatron and LHC through
 the annihilation of a pair of up quarks.
Clearly, a proton-proton collider leads to a higher production rate
for $\overline{\Delta}_{u^cu^c}$ compared to the proton-anti-proton
colliding machine. As a signature of diquark productions at hadron
colliders,
 we concentrate on its decay channel which includes at least
 one anti-top quark (top quark for anti-diquark Higgs case)
 in the final state.
Top quark has large mass and decays electroweakly before
hadronizing. Due to this characteristic feature distinguishable
from other quarks,
 top quarks can be used as an ideal tool \cite{TopPhys}
 to probe other new physics beyond the standard model
 \cite{tp}.

Since diquark couples with only up-type quarks,
 once it is produced, its decay give rise to
 production of double top quarks
 ($\overline{\Delta}_{u^cu^c} \rightarrow tt$)
 and a single top quark + jet
 ($\overline{\Delta}_{u^cu^c} \rightarrow t u$ or $t c$).
These processes have no standard model counterpart,
 and the signature of diquark production would be
 cleanly distinguished from the standard model background.
We leave detailed collider studies on
 signal event of diquark (anti-diquark) Higgs production
 and the standard model background event
 for future works.
Instead, as a conservative treatment,
 we calculate resonant production of diquark
 and anti-diquark Higgs at Tevatron and LHC
 and compare it to $t \overline{t}$ production in the standard model.
The reason is that to observe resonant production of
$\Delta_{u^cu^c}$ and measure its mass,
 it is necessary to reconstruct the invariant mass of
 the final state.
In the double top quark production, if one uses
 the leptonic decay mode of a top quark,
 $t \rightarrow b W^+ \rightarrow b \ell^+ \nu$,
 for the identification of top quark,
 with one missing neutrino and
 the hadronic decay mode for the other top quark
 to reconstruct the invariant mass \cite{topinvmass},
 it becomes difficult to tell $t$ from $\overline{t}$.
However note that if one can use leptonic decay modes
 for both tops, one can distinguish $tt$ from $t\overline{t}$
 through charges of produced leptons.

First, we give basic formulas for our study on diquark Higgs
 production at hadron colliders.
The fundamental processes in question are
 $u u \rightarrow \overline{\Delta}_{u^cu^c} \rightarrow tt, tu, tc$
( $\overline{u} \overline{u} \rightarrow \Delta_{u^cu^c}
 \rightarrow \overline{t} \overline{t}, \overline{u} \overline{t}, \overline{c} \overline{t}$
  for anti-diquark Higgs production).
 From Eq.~(\ref{Yukawa}), the cross section is found to be
\bea && \frac{ d \sigma( uu \rightarrow\overline{ \Delta}_{u^cu^c}
\rightarrow tt)}
 {d \cos \theta}
 = \frac{|f_{11}|^2 \; |f_{33}|^2}{16 \pi} \nonumber\\
&\times &
 \frac{\hat{s}-2 m_t^2}
   {(\hat{s}-m_{\Delta}^2)^2
  + m_{\Delta}^2 \Gamma_{\rm tot}^2 }
   \sqrt{1- \frac{4 m_t^2}{\hat{s}}},  \nonumber  \\
&& \frac{d \sigma( uu \rightarrow \overline{\Delta}_{u^cu^c}
\rightarrow ut, ct)}{d \cos \theta}
 = \frac{|f_{11}|^2 \; |f_{13,23}|^2}{8 \pi \hat{s}} \nonumber\\
&\times &
  \frac{(\hat{s}- m_t^2)^2}
    {(\hat{s}-m_{\Delta}^2)^2
  + m_{\Delta}^2 \Gamma_{\rm tot}^2 } .
\label{CrossParton}
\eea
Here, we have neglected all quark masses
except for top quark mass ($m_t$), $\cos \theta$ is the scattering
angle, and $\Gamma_{\rm tot}$ is the total decay width of diquark
Higgs,  which is the sum of each partial decay width,
\bea
 \Gamma (\overline{\Delta}_{u^cu^c} \to uu, cc)
 &=& \frac{3}{16 \pi}  |f_{11,22}|^2 \; m_{\Delta},   \nonumber \\
 \Gamma (\overline{\Delta}_{u^cu^c} \to tt)
 &=& \frac{3}{16 \pi} |f_{33}|^2 \; m_{\Delta} \nonumber \\
& \times &
  \sqrt{1-\frac{4 m_t^2}{m_{\Delta}^2}}
  \left(  1- \frac{2 m_t^2}{m_{\Delta}^2}    \right)  \nonumber \\
 \Gamma (\overline{\Delta}_{u^cu^c} \to uc)
 &=& \frac{3}{8 \pi} |f_{12}|^2 \; m_{\Delta},  \nonumber \\
 \Gamma (\overline{\Delta}_{u^cu^c} \to u t, c t)
 &=& \frac{3}{8 \pi} |f_{13,23}|^2 \; m_{\Delta}
  \left(  1- \frac{m_t^2}{m_{\Delta}^2} \right)^2.
\eea
Note that the cross section is independent of the scattering angle
 because the diquark Higgs is a scalar.

With these cross sections at the parton level,
 we study the diquark production at Tevatron and LHC.
At Tevatron, the total production cross section
 of an up-type quark pair ($u_i u_j$ where $u_{1,2,3}=u,c,t$)
 through diquark Higgs in the s-channel is given by
\bea
 && \sigma (p \overline{p} \to u_i u_j)
 = \int dx_1 \int dx_2 \int d \cos \theta \nonumber \\
 &\times & f_u(x_1, Q^2)  f_{\overline{u}}(x_2, Q^2)   \nonumber \\
 &\times &
  \frac{d \sigma(u u \to \overline\Delta_{u^cu^c} \to u_i u_j;
  \hat{s}=x_1 x_2 E_{\rm CMS}^2)}{d \cos \theta},
\label{CrossTevatron} \eea where $f_u(x_1, Q^2)$ and $
f_{\overline{u}}(x_2, Q^2)$ denote the parton distribution function,
 and $E_{\rm CMS}$ is the collider energy.
Note that one parton distribution function is for up quark
 and the other is for the sea up quark, since it comes from an
anti-proton
( for a proton-anti-proton system such as at Tevatron).
This fact indicates that at Tevatron
 the production cross section of diquark Higgs is the same
 as the one of anti-diquark Higgs,
 reflecting that the total baryon number of
 initial $p \overline{p}$ state is zero.

At LHC, the total production cross section
 of an up-type quark pair is given by
\bea
&& \sigma (p p \to u_i u_j)
 = \int dx_1 \int dx_2 \int d \cos \theta \nonumber \\
&\times&
   f_u(x_1, Q^2)  f_u(x_2, Q^2)   \nonumber \\
&\times &
  \frac{d \sigma(u u \to \overline\Delta_{u^cu^c} \to u_i u_j;
  \hat{s}=x_1 x_2 E_{\rm CMS}^2)}{d \cos \theta}.
\eea
Here, both of parton distribution functions are for up quark in proton
(both valence quarks),
 corresponding to a proton-proton system at LHC.
Total production cross section
 of an up-type anti-quark pair ($\overline{u}_i \overline{u}_j$)
 is obtained by replacing the parton distribution function
 into the one for anti-quark.
The initial $p p$ state has a positive baryon number,
 so that the production cross section of diquark Higgs
 is much larger than the one of anti-diquark Higgs at LHC.
The dependence of the cross section on
 the final state invariant mass $M_{u_i u_j}$
 is described as
\bea
&& \frac{d \sigma (p p \to u_i u_j)}{d M_{u_i u_j}}
= \int d \cos \theta
 \int^1_{ \frac{M_{u_i u_j}^2}{E_{\rm CMS}^2}} dx_1
 \frac{2 M_{u_iu_j}}{x_1 E_{\rm CMS}^2} \nonumber\\
&\times&  f_u(x_1, Q^2)
  f_u \left( \frac{M_{u_iu_j}^2}{x_1 E_{\rm CMS}^2}, Q^2
 \right)   \nonumber \\
 &\times &
  \frac{d \sigma(u u \to \overline\Delta_{u^cu^c} \to u_i u_j)}
 {d \cos \theta}.
\label{CrossLHC}
\eea

The production cross section of the diquark Higgs and its branching
 ratio to final state up-type quarks depends on the coupling $f_{ij}$.
This coupling is, in general, a free parameter in the model,
 and in our following analysis, we take an example for $f_{ij}$,
%
\bea
 f_{ij}=
\left[
 \begin{array}{ccc}
       0.3 & 0 & 0.3 \\
         0 & 0 & 0   \\
       0.3 & 0 & 0.3
\end{array}
\right] .
\eea
In this example, the phenomenological constraints on $f_{ij}$
 discussed in the previous section are satisfied
 with $f_{12}=f_{22}=0$.
This example gives rise to processes,
 $ uu \to tt, ut$, that we are interested in.

Let us first examine the lower bound on the diquark Higgs mass
 from Tevatron experiments.
We refer the current experimental data of
 the cross section of top quark pair production \cite{CDF},
\bea \sigma(t \overline{t}) = 7.3
  \pm 0.5({\rm stat})  \pm 0.6({\rm syst})  \pm 0.4({\rm lum})
  \; {\rm pb},
\eea
 and impose a constraint for the double top quark and a single top quark
 production cross sections through diquark Higgs in the s-channel.
 Since most of the $\sigma_{t\overline{t}}$ value can be
 understood as the standard model effect, the possible new physics
 should be in the uncertainty range of $\sigma_{t\overline
 t}$, we take the following conservative bound as
\bea \sigma( p \overline{p} \to \Delta_{u^cu^c} \to tt, ut)
 \lesssim 1.5 {\rm pb}.
\eea In our numerical analysis, we employ CTEQ5M \cite{CTEQ}
 for the parton distribution functions
 with the factorization scale $Q=m_t=172$ GeV.
Fig.~1 shows the total cross section of $tt$ and $tu$ productions
 as a function of the diquark Higgs mass,
 with $E_{\rm CMS} = 1.98$ TeV.
The lower bound is found to be $m_{\Delta} \gtrsim$ 470 GeV.

\begin{figure}[t]
\includegraphics[width=0.8\columnwidth]{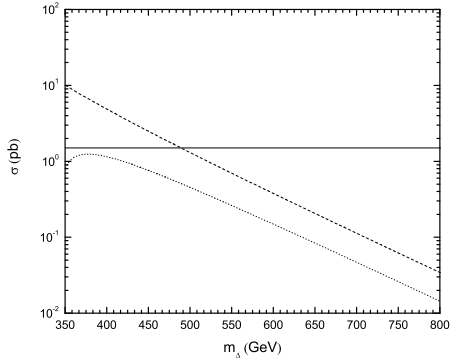}
\caption{ The cross sections of $tt$ (dotted line)
 and $tj$ (dashed line) productions
 mediated by the diquark Higgs in s-channel at Tevatron
 with $E_{\rm CMS}=1.96$ TeV.
}
\end{figure}

\begin{figure}[t]
\includegraphics[width=0.8\columnwidth]{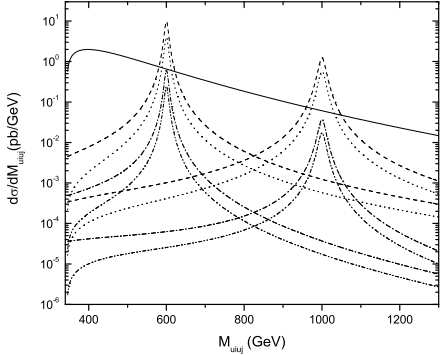}
\caption{
 The differential cross sections for $tj$ (dashed line), $tt$
 (dotted line), $\overline{t}j$ (dashed-dotted line) and
 $\overline{t}\overline{t}$ (dashed-dotted-dotted line)
 as a function of the invariant mass of final state $M_{u_iu_j}$.
 The left peak corresponds to $m_\Delta=600(\rm GeV)$
 and the right one to $m_\Delta=1$ TeV.
 The solid line is the standard model $t\overline{t}$ background.
}
\label{Fig2}
\end{figure}

\begin{figure}[t]
\includegraphics[width=0.8\columnwidth]{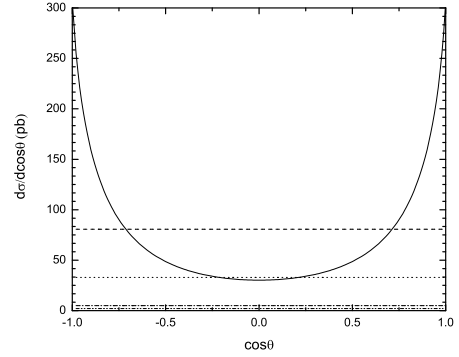}
\caption{
Angular distribution of the cross section
 for $m_{\Delta}=600$ GeV
 with $M_{\rm cut}=550$ GeV,
 together with the $t \overline{t}$ production in the standard model.
 The same line convention as in the Fig.~\ref{Fig2}
 has been used.
}\label{Fig3}
\end{figure}

Next we investigate the diquark and anti-diquark Higgs production
 at LHC with $E_{\rm CMS}=14$ TeV.
The differential cross sections for each process
 with $m_{\Delta}=600$ GeV and 1 TeV are depicted in Fig.~2,
 together with the $t \overline{t}$ production cross section
 in the standard model.
We can see that
 the peak cross sections for the $tt$ and $tu$ productions
 exceed the standard model cross section
 while the $\overline{t} \overline{t}$ and $\overline{t} \overline{u}$
 cross sections are lower than it.
This discrepancy between the production cross sections
 of diquark and anti-diquark Higgs at LHC is the direct evidence
 of the non-zero baryon number of diquark Higgs.
The charge of the lepton from leptonic decay of top quark
 or anti-top quark can distinguish top quark from anti-top quark.
Counting the number of top quark events and anti-top quark events
 from their leptonic decay modes would reveal
 non-zero baryon number of diquark Higgs.

The angular distribution of the final states carries
 the information of the spin of the intermediate states.
As shown in Eq.~(\ref{CrossParton}),
 there is no angular dependence on the diquark Higgs production
 cross section, because the diquark Higgs is a scalar particle.
On the other hand, the top quark pair production in the standard model
 is dominated by the gluon fusion process,
 and the differential cross section shows peaks
 in the forward and backward region.
Therefore, the signal of the diquark Higgs production
 is enhanced at the region with a large scattering angle
 (in center of mass frame of colliding partons).
Imposing a lower cut on the invariant mass $M_{\rm cut}$,
 the angular dependence of the cross section is described as
\bea
&& \frac{d \sigma (p p \to u_i u_j) } {d \cos \theta}
= \int_{M_{\rm cut}}^{E_{\rm CMS}}
 d M_{u_i u_j}
 \int^1_{ \frac{M_{u_i u_j}^2}{E_{\rm CMS}^2} } dx_1    \nonumber \\
&\times&
 \frac{2 M_{u_iu_j}}{x_1 E_{\rm CMS}^2}
 f_u(x_1, Q^2) f_u
\left( \frac{M_{u_iu_j}^2}{x_1 E_{\rm CMS}^2}, Q^2
 \right)
  \nonumber \\
 &\times &
  \frac{d \sigma(u u \to \Delta_{u^cu^c} \to u_i u_j)}
 {d \cos \theta}.
\label{DCrossLHC}
\eea
The results for $m_{\Delta}=600$ GeV
 with $M_{\rm cut}=550$ GeV are depicted in Fig.~3,
 together with the standard model result.
Here the lower cut on the invariant mass
 close to the diquark Higgs mass
 dramatically reduces the standard model cross section
 compared to the diquark Higgs signal.

We now discuss the connection of the coupling $f_{ij}$ to the
neutrino mass. Once the $B-L$ symmetry is broken by $\langle
\Delta^c \rangle$ along the $\nu^c\nu^c$ direction,
 right-handed neutrinos acquire masses
 through the Yukawa coupling in Eq.~(\ref{Yukawa})
 and their mass matrix is proportional to $f_{ij}$.
Therefore, $f_{ij}$ is related to neutrino oscillation data
 though the (type I) see-saw mechanism which unfortunately
 involves unknown Dirac Yukawa couplings.
When we impose the left-right symmetry on a model,
 $\Delta^c$ is accompanied by
 $\overline{\Delta} : ({\bf 3,1},\overline{\bf 10})$,
 which adds a new term to the superpotential $ f \psi \overline{\Delta}^c_L \psi$
 with the same Yukawa coupling $f_{ij}$.
Through this Yukawa coupling, the type II see-saw mechanism can
 generates Majorana masses for left handed neutrinos.
When the type II see-saw contributions dominate
 the light neutrino mass matrix becomes proportional to $f_{ij}$.
In this case, there is a direct relation between the collider
physics involving diquark Higgs production and neutrino
oscillation data.

For the type II see-saw dominance, a sample value for $f_{ij}$
that fits neutrino observations is given by,
\bea
 f_{ij}=
\left[
 \begin{array}{ccc}
       0.27  &  -0.48 & -0.47 \\
       -0.48 &     0  & -0.38 \\
       -0.47 &  -0.38 &  0.2
 \end{array}
\right] . \eea Again, this Yukawa coupling matrix is consistent
with phenomenological
 constraints discussed in the previous section.
The type II see-saw gives the light neutrino mass matrix
 via $m_\nu = f v_T$
 with 
 $v_T= \langle \overline{\Delta} \rangle$.
For $v_T=0.1$ eV, it predicts neutrino oscillation parameters to be:
\bea
 && \Delta m_{12}^2 = 8.9 \times 10^{-5} \; {\rm eV}^2,
 \; \; \Delta m_{23}^2 = 3 \times 10^{-3} \; {\rm eV}^2,
\nonumber \\
 && \sin^2  \theta_{12} = 0.32, \; \;
 \sin^2  2 \theta_{23} = 0.99,  \; \;
 |U_{e3}| = 0.2,   \nonumber
\eea
which are all consistent with the current
 neutrino oscillation data \cite{PDG}.
Here the resultant light neutrino mass spectrum
 is the inverse hierarchical.
For $f_{22} \ll1 $ as required by $D^0-\overline{D^0}$ mixing
data,
 analytic and numerical studies show that
 only the inverse hierarchical mass spectrum can
 reproduce the observed neutrino oscillation data for the type II
 seesaw case.

We have performed the same analysis as before for this case and
find the lower bound on the diquark Higgs mass from Tevatron data
 to be $m_{\Delta} \gtrsim$ 450 GeV,
 which is a little milder than before.
In this case, the peak cross section of only the single
 top + jet production exceeds the $t \overline{t}$ production cross section
 of the standard model.
The differential cross section of Eq.~(\ref{DCrossLHC})
 is independent of the scattering angle,
 and we find $d \sigma/d \cos \theta = 60.6$ pb
 for the single top + jet production
 for $m_{\Delta}=600$ GeV with $M_{\rm cut}=550$ GeV.

Finally, we comment on spin polarization of
 the final state top (anti-top) quark.
Because of its large mass, top quark decays
 before hadronizing and the information of
 the top quark spin polarization is directly transferred
 to its decay products and results in
 significant angular correlations between the top quark
 polarization axes and the direction of motion
 of the decay products \cite{TopSpin}.
Measuring the top spin polarization provides
 the information on the chirality nature of top quark
 in its interaction vertex.
It has been shown that measuring top spin correlations
 can increase the sensitivity to a new particle
 at Tevatron \cite{SpinCorr1} and LHC \cite{SpinCorr2}.
In the diquark Higgs production, it is very interesting
 to measure the polarization of top (anti-top) quark
 in the single top production.
Only the right-handed top quark couples to diquark Higgs
 and the top quark produced from diquark Higgs decay
 is right-handed state, while top quark from the single top production
 through electroweak processes in the standard model
 is purely left-handed.

\acknowledgments
We like to thank K. S. Babu, T. Han, K. Smolek and C. P. Yuan for
useful discussions. The works of R.N.M. and H.B.Y. are supported by
the National Science Foundation Grant No. PHY-0652363. The work of
N.O. is supported in part by the Grant-in-Aid
 for Scientific Research from the Ministry of Education,
 Science and Culture of Japan (\#18740170).



\end{document}